\documentclass[aps,pra,floatfix,nofootinbib,showpacs]{revtex4} 
\usepackage{graphicx} 
\usepackage{amsmath} 
\usepackage{amssymb}
\usepackage{subfigure}
\newcommand{\BEQ}{\begin{equation}}
\newcommand{\EEQ}{\end{equation}}
\newcommand{\BEA}{\begin{eqnarray}}
\newcommand{\EEA}{\end{eqnarray}}

\renewcommand{\d}{{\rm d}}

\newcommand{\ssum}{{\sum}}
\newcommand{\pprod}{{\prod}}

\renewcommand{\a}{\alpha}

\newcommand{\g}{\gamma}
\newcommand{\tc}{{\tilde c}}
\newcommand{\tx}{{\tilde x}}

\newcommand{\bJ}{\bar{J}}
\newcommand{\bars}{\bar{s}}
\newcommand{\bs}{{ \bf s}}

\newcommand{\at}{{\rm atanh}}

\newcommand{\bA}{{\bf A}}

\newcommand{\bp}{{\bf p}}

\newcommand{\comment}[1]{}

\begin{document}

\title {Statistical Mechanics of Semi--Supervised Clustering in Sparse Graphs}

\author{ Greg Ver Steeg$^1$, Aram Galstyan$^1$, and Armen E. Allahverdyan$^2$}
\address{
$^1$Information Sciences Institute, University of Southern California,
Marina del Rey, CA 90292, USA\\
$^2$Yerevan Physics Institute,
Alikhanian Brothers Street 2, Yerevan 375036, Armenia}

\begin{abstract} 
We theoretically study semi--supervised clustering in sparse graphs in the presence of pair--wise constraints on the cluster assignments of nodes.  We focus on bi--cluster graphs, and study the impact of semi--supervision for varying constraint density and overlap between the clusters.  Recent results for unsupervised clustering in sparse graphs  indicate that there is a critical ratio of within--cluster and between--cluster connectivities below which clusters cannot be recovered with better than random accuracy. The goal of this paper is to examine the impact of pair--wise constraints on the clustering accuracy.  
Our results suggests that when the density of the constraints is sufficiently small, their only impact is to shift the detection threshold while preserving the criticality. Conversely, if the density of (hard) constraints is above the percolation threshold, the criticality is suppressed and the detection threshold disappears. 
\end{abstract}


\maketitle
\section{Introduction}
Most real--world networks have a modular structure, i.e., they are composed of {\em clusters} of  well connected nodes, with relatively lower density of links  across different clusters~\cite{newman1,Newman2003,M.E.J.Newman06062006,NewmanGirvan2004}. Those modules  might represent groups of individuals with shared interests in online social networks, functionally coherent units  in protein--protein interaction networks, or topic--specific research communities in co--authorship networks. Modularity is important for understanding various structural and dynamical properties of networks~\cite{arenas2006,Galstyan2007,Gleeson2008,Pan2007,Dorogovstev2008,Ostilli2009}. Consequently, the problem of detecting  modules, or clusters, has  recently attracted a considerable interest both in statistical physics and computer science communities  (for a recent comprehensive review of existing approaches see~\cite{Fortunato2010}). 

Along with algorithmic development,  recent research has focused on characterizing statistical significance of clusters detected by different methods~\cite{Guimera2004,Fortunato2007,Karrer2008,Lancichinetti2010,Good2010}. Another fundamental issue is the {\em feasibility} of detecting clusters, assuming they are present in the network. To be specific, consider the so called {\em planted bi--partition} model~\cite{karp}, which is a special case of more general family of generative models known as  {\em stochastic block-models}~\cite{Holland1983,wang,Nowicki2001}. In this model the nodes are partitioned into two equal--sized groups, and each link within a group and between the groups is present with probabilities $p$ and $r$, respectively.  An important question is  how well one can recover the latent cluster structure  in the limit of large network sizes. It is known that in dense  networks where  the average connectivity scales linearly with the number of nodes $N$ (e.g., $p$ and $r$ are constants), the clusters in the  planted partition model  can be recovered with  asymptotically {\em perfect} accuracy  for any  $p-r> N^{-1/2+\epsilon}$~\cite{karp}. Recently, a more general result  obtained for a larger class of stochastic block--models states that  certain statistical inference methods are asymptotically consistent provided that the average connectivity grows faster than $\log N$~\cite{Bickel2009}.

The situation is  significantly  different for sparse graphs, where the average connectivity remains finite as in the asymptotic limit $N\rightarrow \infty$. Indeed,  recently it was shown~\cite{leone} that   planted partition models (of arbitrary topology) with finite connectivities are characterized by a phase transition from {\em detectable } to  {\em undetectable} regimes as one increases the overlap between the clusters, with the transition point depending on the actual degree distribution of the partitions. In particular, let $p=\a/N$, $r=\g/N$, where $\a$ and $\g$ are finite average connectivities {\em within} and {\em across} the clusters, and  let $p_{in}=\a/(\a+\g)$ be the fraction of links that fall within the clusters, so that $p_{in}=1$ and $p_{in} = \frac{1}{2}$ correspond to perfectly separated and perfectly mixed clusters, respectively.  Then  there is a critical value  $p_{in}^c=\frac{1}{2}+\Delta$, $\Delta>0$ such that for $p_{in}<p_{in}^c$ the clusters cannot be recovered with {\em better than random accuracy} in the asymptotic limit. When the planted clusters have Erdos--Renyi topology,  one can show that $\Delta \propto 1/\sqrt{\a + \g}$ for large $(\a+\g)$~\cite{armen}.

From the perspective of statistical inference, this type of phase transition between detectable and undetectable regimes is undesirable, as it signals inference instabilities -- large fluctuations in accuracy in response to relatively small shifts in the parameters.  In~\cite{armen} it was shown that this instability can be avoided if one uses prior knowledge about the underlying group structure. Namely, it was demonstrated that knowing the correct cluster assignments for arbitrarily small but {\em finite} fraction of nodes destroys the criticality and moves the detection threshold to its intuitive (dense--network limit) value $p_{in}=\frac{1}{2}$, or $\a=\g$. This can be viewed as a {\em semi--supervised} version of the problem, as opposed to an unsupervised version where the only available information is the observed graph structure. 

In practice, semi--supervised graph clustering methods can utilize two types of  background knowledge -- cluster  assignment for a subset of nodes~\cite{zhu,getz}, or pair--wise constraints in the form of {\em must-link} ({\em cannot-links}), which imply that a pair of nodes must  be (cannot be) assigned to the same group~\cite{basu2004,kulis}. In fact, the latter type of constraints are more realistic in scenarios where it is easier to assess  similarity of two objects rather than to label them individually. The goal of this paper is to examine the impact of latter form of semi-supervision on clustering accuracy.   
Specifically, we focus on a random network composed of two equal--sized clusters,  where the clustering objective can mapped to an appropriately defined Ising model defined on the planted partition graph.

The rest of the paper is organized as follow. In the next section we reduce the  block--structure  estimation problem  to an appropriately defined Ising model, and describe  the zero temperature cavity   method used to study the properties of the model. In Section~\ref{sec:M2} we analytically study a specific case of soft constraints. In Section~\ref{sec:hard} we consider the case of hard constraints, and study it using population dynamics methods. We finish the paper by discussing our main results in Section~\ref{sec:discussion}.

\section{Model}
\label{sec:model}
Consider a graph with two clusters containing $N$--nodes each. Each pair of nodes within the same cluster is linked with probability $p=\a/N$, where $\a$ is the average within--cluster connectivity. Also, each pair of nodes in different clusters is linked with probability $r=\g/N$, where $\g$ is inter--cluster connectivity. Let $s_i=\pm 1$ denote the cluster assignment of node $i$, and let $\bs=(s_1,\dots,s_{2N})$ denote a cluster assignment for all the nodes in the network.  Further, let  $\bA$ be the observed adjacency matrix of interaction graph of  $2N$ nodes so that $A_{ij}=1$ if we have observed a link between nodes $i$ and $j$, and $A_{ij}=0$ otherwise. Within the  above model, the conditional distribution of observation for a given configuration $\bs$ reads
\BEA
\bp(\bA| \bs) = p^{c_{+}}[1-p]^{c_{-}}r^{d_{+}}[1-r]^{d_{-}}
\EEA

Here  $c_+$, $d_+$ ($c_-$, $d_-$) are the total number of observed (missing) links within and across the groups,
\BEA
c_+=\ssum_{i<j}A_{ij}\delta_{s_i, s_j}\ ,\ c_- = N(N-1) - c_+\\
d_+=\ssum_{i<j}A_{ij}(1-\delta_{s_i, s_j})\ ,\ d_-  = N^2-d_+,  
\EEA
where $\delta_{ij} =1$ if $i=j$ and $\delta_{ij} =0$ otherwise.  Let us define $J_{NL}=\ln[(1-r)/(1-p)]$, $J_{L}=\ln[p/r]+J_{NL}$. Then  the log of the joint distribution over both observed and hidden variables can be written as follows~\cite{Hastings2006,Hofman2008} : 
\BEA
H(\bs, \bA) = -\ln [\bp(\bA| \bs)\bp(\bs)] = -\ssum_{i>j}J_{L}A_{ij}\delta_{s_i,s_j}  + \ssum_{i>j}J_{NL}\delta_{s_i,s_j}+H_{\pi}({\bs})
\label{eq:hamnet}
\EEA
Here $J_L$  and $J_{NL}$ stand for the contributions from observed links and non--links, respectively, while the last term $H_{\pi}(\bs)$ encodes prior information about the latent structure one might have.  In the scenario considered below, we assume that the cluster sizes are known a priori. This constraint can be forced by the appropriate choice of $H_{\pi}(\bs)$, so that any clustering arrangement that violates the size constraint will be disallowed. Then it is easy to check that the second term amounts to a constant that can be ignored. Furthermore,  since below we are interested in the minimum of $H(\bs, \bA)$, we can set $J_{L}=1$ without loss of generality.


We find it convenient to make  the bi--cluster nature of the network explicit by introducing separate  variables $s_i=\pm 1$ and $\bars_i=\pm 1$
($i=1,\ldots,N$) for two groups. Then Eq.~\ref{eq:hamnet} is reduced to the following Ising Hamiltonian (aside from an unessential  scaling factor): 
\BEA
\label{ham}
H =- \sum_{i<j}^N J_{ij}s_is_j - \sum_{i<j}^N\bJ_{ij}\bars_i\bars_j- \sum_{i,j}^NK_{ij}s_i\bars_j + H_{\pi}(\bs) \ . 
\EEA
Here $J_{ij}$ and $\bJ_{ij}$ are the elements on two diagonal blocks  of the matrix $\bA$ describing the connectivity within each cluster, whereas $K_{ij}$-s are the elements on the (upper) off--diagonal block of $\bA$ that link nodes across the clusters. In the unsupervised block--model,  they are random Bernoulli trials with parameters $p$ and $r$. 

To account for background information in the form of pairwise constraints, we use the following form for the prior part of the Hamiltonian: 
\BEA
\label{prior}
H_{\pi}(\bs) =- w_{m}\sum_{i<j}^N [ \theta _{ij}s_is_j + { \bar \theta}_{ij}\bars_i\bars_j] + w_{c}  \sum_{i,j}^N\phi_{ij}s_i\bars_j  \ . 
\EEA
where $\theta _{ij}=1$ (${ \bar \theta}_{ij}=1$) if the corresponding pair of nodes are connected via a must--link  constraint within the first (second) cluster, and   $\theta _{ij}=0$ (${ \bar \theta}_{ij}=0$) otherwise. Similarly, $\phi _{ij}=1$ if there is a cannot link between corresponding nodes in respective clusters, and   $\phi _{ij}=0$ if there is no such link. Here $w_{m}$ and $w_{c}$ are the costs of violating a must--link and cannot--link constraints, respectively.  For the sake of simplicity, below we will choose $w_{m}=w_{c} \equiv w$, where $w$ is a positive integer. 

 Below we will assume that the constraints are introduced randomly and independently for each pair of nodes. Namely,  $\theta _{ij}, { \bar \theta}_{ij}$-s and $\phi_{ij}$  are Bernoulli trials  with parameter $f_{m}$ and $f_{c}$, respectively. Then the prior part of the Hamiltonian can be absorbed into~\ref{ham} by the following choice for the distribution of the couplings  in~\ref{ham}:   
\BEA
p(J_{ij}) &=& [1- p]\delta(J_{ij}) +   p[1-f_{m}] \delta(J_{ij}-1) +   pf_{m} \delta(J_{ij}-w)\\
p(K_{ij}) &=& [1- r ]\delta(K_{ij}) +   r[1-f_{c}] \delta(K_{ij}-1) +   rf_{c} \delta(K_{ij}+w)
\label{JK}
\EEA

We are interested in the properties of the above Hamiltonian~\ref{ham} in the limit of large $N$. 
Below we study it within the  Bethe--Peierls approximation.  Let $P(h)$ $(\bar{P}(h)$) denote the probability of an internal ({\em cavity}) field acting on an $s$ ($\bars$) spin. Then we have according to the zero temperature cavity method~\cite{mezard3,mezard4}: 
\BEA
\label{cavity}
P(h)&=&
\int \pprod_{n=1}^N dJ_{0n}p(J_{0n}) \int \pprod_{n=1}^N dK_{0n}p(K_{0n}) 
\int \pprod_{k=1}^N P(h_k)\d h_k\pprod_{l=1}^N \bar{P}(g_k)\d g_k
\nonumber\\
&\times& \delta\left[h-\ssum_{k=1}^N \phi[h_k,J_{0k}] - \ssum_{k=1}^N \phi[g_k,K_{0k}] \right],
\EEA
where $\phi[a,b]\equiv{\rm sign}(a){\rm sign}(b)\,{\rm min}[\, |a|, |b|\,] $, and where
$g_k$ (resp. $h_k$) are the fields acting on the $s$-spin from
$\bars$-spin (respectively from other $s$-spins). 

 Using the integral representation of the delta function, performing the integration over the coupling constants, and employing the symmetry $\bar{P}(g)=P(-g)$, we obtain in the limit $N \rightarrow \infty$
\BEA
\label{cavity1}
 P(h)&=& \int \frac{dz}{2 \pi} e^{izh} e^{-\a -\g  } \\
 &\times&\exp  \left \{  \int dg P(g) \left [ \a (1-f_{m})  e^{-iz\,{\rm sign}(g)\,{\rm min}[|g|, 1]} +    \g (1-f_{c})e^{iz\,{\rm sign}(g)\,{\rm min}[|g|, 1]} + [\a f_{m} + \g f_{c}]  e^{-iz\,{\rm sign}(g)\,{\rm min}[|g|, w]} \right ] \right \}  \nonumber
 \EEA
Once $P(h)$ is found we obtain the first two moments of $s_i$ as 
\BEA
\label{arto}
m=\int P(h) \, {\rm sign}(h), \qquad q=\int P(h) \, {\rm sign}^2(h), \ldots,
\EEA
Here $m$ is the magnetization averaged over   the graph structure (including the constraints)
(i.e., averaging over $J_{ij}$, $\bJ_{ij}$ and $K_{ij}$), and Gibbs distribution, which, at zero temperature case considered here it means
averaging over all configurations of $s_i$ and ${\bar s}_i$ that in the thermodynamic limit have| the same (minimal) values of the Hamiltonian $H$.  And  $q$ is called 
Edwards-Anderson (EA) order parameter; $q$ differs from $1$ due to possible
contribution $\propto \delta(h)$ in $P(h)$. Note that the accuracy of the clustering (i.e., probability that a node has been assigned to the correct cluster) is simply $\frac{1+|m|}{2}$. Thus, $|m|=1$ corresponds to perfect clustering, whereas $m=0$ means that discovered clusters have only random overlap with the true cluster assignments. 


In the following, we assume that the constraints are generated with uniform probability $f_m=f_c=\rho/(\a+\g)$, where $\rho$ is the average number of constraints per node. 

\section{Soft Constraints}
\label{sec:M2}

We first consider the case when violating a constraint carries a finite cost. The most trivial such case is when $w=1$. In this case, the must--link constraints do not yield any additional information. The cannot-link constraints, however, help clustering by ``flipping" the sign of the corresponding edge, thus favoring anti--ferromagnetic interactions between the nodes across different clusters. In fact, it can be shown from Eq.~\ref{cavity1} that the only impact of the constraints with $w=1$ is to renormalize within and across cluster connectivities, $(\a,\g) \rightarrow (\a + \rho \g, \g - \rho\g)$. Recall that the {\em mixing} parameter is defined as $p_{in}=\frac{\a}{\a+\g}$. Thus, this situation is identical to the un--supervised clustering scenario~\cite{leone,armen} with renormalized mixing parameter $p_{in} \rightarrow  p_{in} + \rho(1-p_{in})/(\a+\g)$. The sole impact of constraints is to shift the detection below which clusters cannot be detected with better than random accuracy. In particular, the modified threshold coincides with its dense network limit $\frac{1}{2}$ for $\rho=\frac{\a+\g}{2}\frac{1-2p_{in}}{1-p_{in}}$.

The role of semi--supervising is qualitatively similar for $w=2$ that we consider next.  Due to the network sparsity, we  search for the solution in the form 
\BEQ
P(h)=\ssum_{-\infty}^{\infty}c_n \delta(h-n) \ ,  \ssum_{-\infty}^{\infty}c_n = 1\ .
\EEQ
 Let us define the following four order--parameters:
\BEA
q = \sum_{k=1}^{\infty}[c_k + c_{-k}] \ , \ m= \sum_{k=1}^{\infty}[c_k -c_{-k}] \ , \  q_1= c_1 +c_{-1} \ , \ m_1= c_1 -c_{-1}
\label{orderparams}
\EEA
Then  Equation~\ref{cavity1} can be represented as follows:
\BEA
\label{cavity3}
P(h)=  e^{-(\a + \g)q}  \int \frac{dz}{2 \pi} e^{izh}
 \prod_{n=1}^{2} e^{\frac{x_n}{2} (e^{y_n + inz} + e^{-y_n- inz})}
\EEA
where
\BEA
x_1 &=&  \sqrt{ [(\a+\g - \rho)q + \rho q_1]^2  - [(\a+\g - \rho)\kappa m + \rho m_1]^2} \\
 y_1 &=& -\at \frac{(\a+\g - \rho)\kappa m +\rho m_1}{(\a +\g - \rho)q + \rho q_1} \\
x_2 &=& \rho  \sqrt{(q - q_1)^2 - (m - m_1)^2} \\
y_2 &=& -\at \frac{m-m_1}{q - q_1}\\
\kappa&=&\frac{\a-\g}{\a+\g}\equiv 2p_{in}-1\ .
\label{x1x2}
\EEA
To proceed further, we express the integrands in the rhs of Eq.~\ref{cavity3}  through modified Bessel functions, and obtain for the coefficients
\BEA
\label{eq:cp}
c_l = e^{-(\a + \g  )q}  \sum_{n=-\infty}^{\infty} I_{2n-l }(x_1) I_{n}(x_2)e^{- y_1(l - 2n) - n y_2 }
\EEA
Furthermore, combining Equations~\ref{eq:cp}  with Eq.~\ref{orderparams} yield the following closed  system for the order parameters of the model:
\BEA
q&=& 2e^{-(\a + \g)q}  \sum_{n=-\infty}^{\infty} \sum_{l=1}^{\infty} I_{2n-l }(x_1) I_{n}(x_2) \cosh [  2ny_1  -y_1l - ny_2  ]\label{sys11} \\
q_1&=& 2e^{-(\a + \g)q}  \sum_{n=-\infty}^{\infty} I_{2n-1 }(x_1) I_{n}(x_2) \cosh [  2ny_1  -y_1 - ny_2  ]  \label{sys22} \\
m &=& 2e^{-(\a + \g)q}  \sum_{n=-\infty}^{\infty} \sum_{l=1}^{\infty} I_{ 2n-l }(x_1) I_{n}(x_2) \sinh [  2ny_1  -y_1l - ny_2  ]\label{sys33}\\
m_1 &=& 2e^{-(\a + \g )q}  \sum_{n=-\infty}^{\infty} I_{2n-1 }(x_1) I_{n}(x_2) \sinh [  2ny_1  -y_1 - ny_2  ] \label{sys44} 
\EEA
We now analyze these equations in more detail.

Again, the case $\rho=0$ corresponds to  the unsupervised  scenario studied in~\cite{leone}, and the system Eqs.~\ref{sys11}, \ref{sys33} predicts a second order phase transitions in parameters $m$ and $m_1$ at a critical mixing parameter value $p_{in}^c$. Below the critical point the magnetization $m$ is zero. Recall that the clustering accuracy (i.e.,  fraction of correct cluster assignments) is given as $\frac{1+|m|}{2}$. Thus,   for any $p_{in}<p_{in}^c$ the estimation cannot do any better than random guessing.  

\begin{figure}[!htb]
\centering
\subfigure[]{
    \includegraphics[width = 0.45\textwidth ]{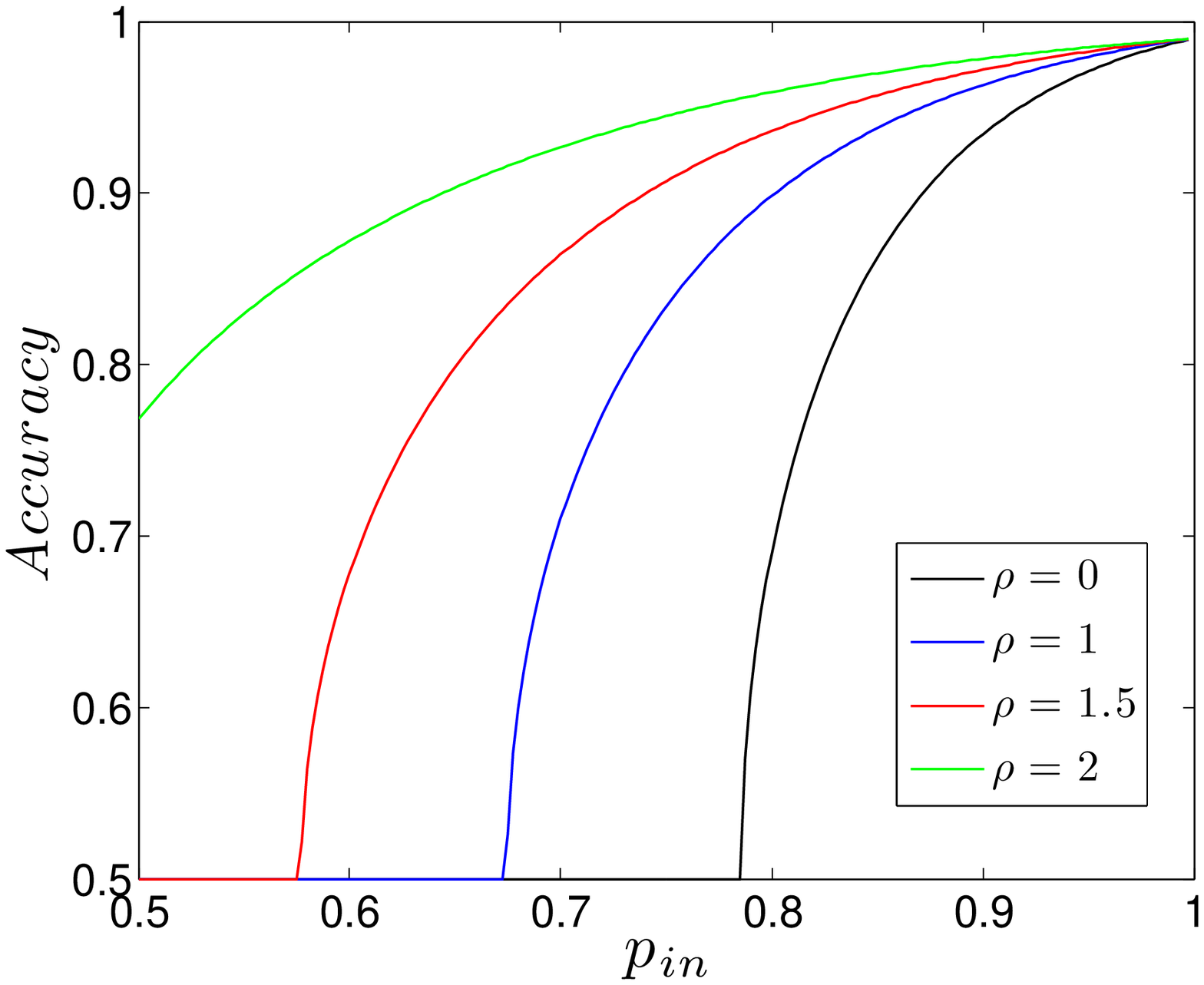} \label{f1a}
    } 
    \subfigure[]{
    \includegraphics[width = 0.45\textwidth]{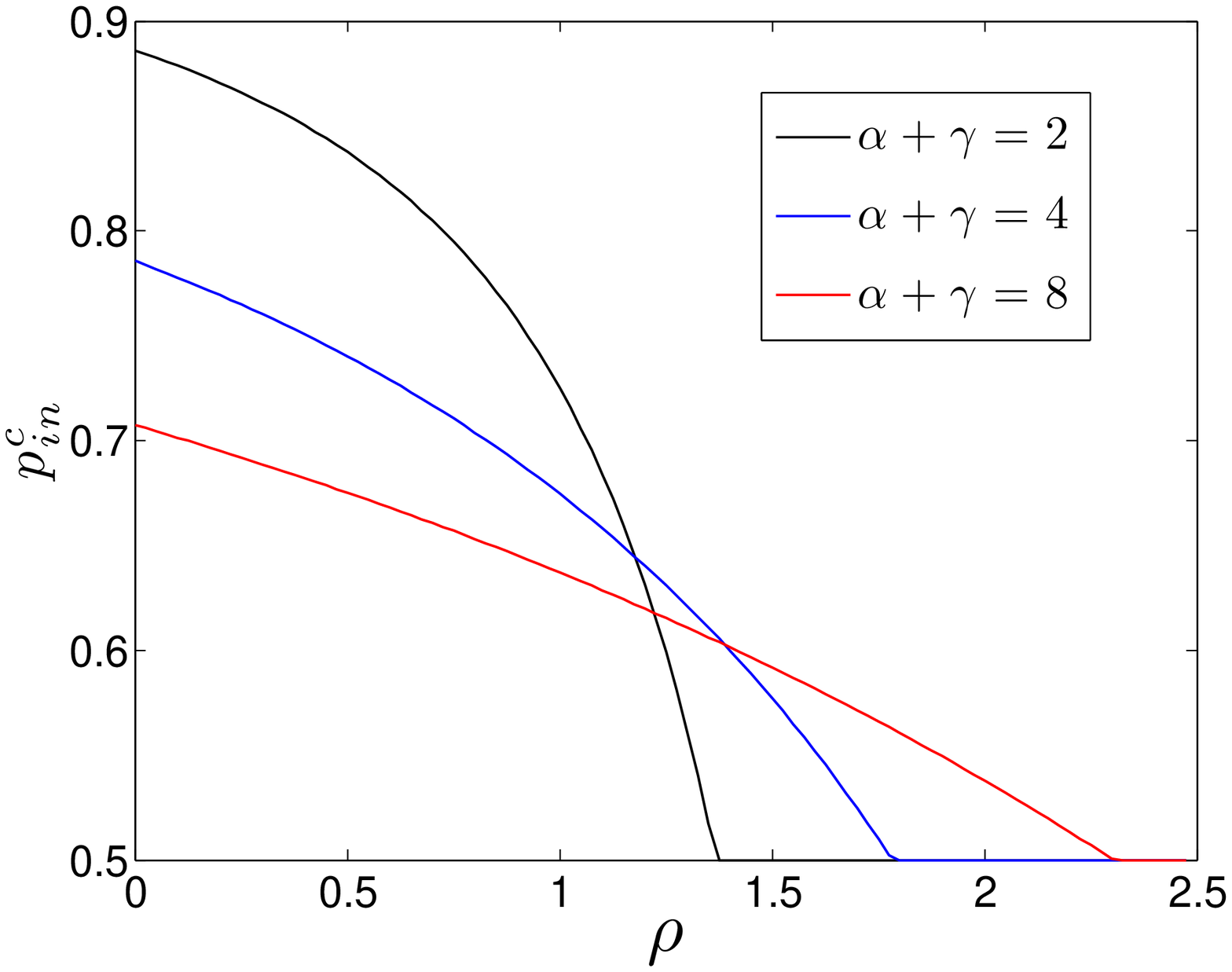} \label{f1b}
    }
   
        \caption{(a) Clustering accuracy $m$ vs. $p_{in}$ for $\a+\g=4$ and different values of $\rho$: From  bottom to top we have $\rho=0, 0.5,1,2$.   (b) Dependence of the critical mixing parameter $p_{in}^c$ on $\rho$. }
\end{figure}

This transition persists also for $\rho>0$:  The presence of the constraints merely shifts the transition point to lower values of $p_{in}^c$.  This is  shown in Figure~\ref{f1a} where we plot the clustering accuracy as a function of $p_{in}$ for $\a+\g=4$. At a certain value of $\rho$, the detection threshold becomes $p_{in}^c=\frac{1}{2}$. If $\rho$ is increased even further, the model has a non--zero magnetization even when $p_{in}=\frac{1}{2}$.  

Note that this shift suggests highly non--linear effect from the added constraints depending on the network parameters. Indeed, when the connectivity is close to its critical value, the constraints can significantly improve the clustering accuracy by moving the system away from the critical regime. And when the system is away from the critical region to start with, then the addition of the constraints might yield no improvement at all compared to the unsupervised scenario. To be more specific, consider Figure~\ref{f1a}, and consider on the impact of semi--supervision on two different network parameters, $p_{in}=0.7$ and $p_{in}=0.65$. We observe that, compared to unsupervised clustering, adding $\rho=1$ constraint per node has drastically different outcomes for those two parameter values. For $p_{in}=0.7$, the addition of the constraint changes the clustering accuracy from $50\%$ (random guessing) to nearly $\sim70\%$. However, adding the same amount of constraints has no impact on accuracy for the network with $p_{in}=0.65$. 

To find the position of the phase transition, we linearize Eqs.~\ref{sys33},~\ref{sys44} around $m=m_1=0$:
\BEA
m&=&-2e^{-(\a+\g)q}\sum_{n=-\infty}^{\infty}\sum_{l=1}^{\infty}I_{2n-l}(\tx_1)I_{n}(\tx_2) \left [ \frac{2n-l}{\tx_1}[(\a+\g-\rho)\kappa m +\rho m_1] -\frac{n}{\tx_2}\rho(m-m_1)\right ] \label{mtrans}\\
m_1&=&-2e^{-(\a+\g)q}\sum_{n=-\infty}^{\infty}I_{2n-1}(\tx_1)I_{n}(\tx_2) \left [ \frac{2n-1}{\tx_1}[(\a+\g-\rho)\kappa m +\rho m_1] -\frac{n}{\tx_2}\rho(m-m_1)\right ] \label{m1trans}. 
\EEA
Here
\BEQ
\tx_1 = (\a + \g - \rho)q + \rho q_1 \ , \ \tx_2 = \rho(q-q_1) \ .
\EEQ
and $q, q_1$ are the solutions of Eqs.~\ref{sys11},~\ref{sys22} with $m=m_1\equiv0$. 

Further simplification of Eqs.~\ref{mtrans},~\ref{m1trans} is carried out by using the the identity $2zI_{k}(z)=k[I_{k-1}(z) - I_{k+1}(z)]$. Let us define
\BEQ
\tc_k = e^{-(\a+\g)q} \sum_{n=-\infty}^{\infty}I_n(\tx_2)I_{2n-k}(\tx_1)
\EEQ
Then the phase transition corresponds to $\det \Lambda=0$ where the elements of the $2\times2$ matrix  $\Lambda$ are as follows:
\BEA
\Lambda_{11}&=&-1+(\tc_0+\tc_1)(\a+\g-\rho)\kappa + \rho(\tc_{0}+2\tc_{1}+\tc_{2})\\
\Lambda_{12}&=&-\rho(\tc_{1}+\tc_{2}) \\
\Lambda_{21}&=&(\tc_0-\tc_2)(\a+\g-\rho)\kappa + \rho(\tc_{1}-\tc_{3})\\
\Lambda_{22}&=&-1+\rho(\tc_0+\tc_3 - \tc_{1}-\tc_{2}) \\
\EEA
Assume a fixed  connectivity $\a+\g$. Then, for each value of $\rho$, there is a critical value of mixing parameter $p_{in}^c$ below which clusters are undetectable. Similarly, for each value $p_{in}$, there is a critical $\rho_c\equiv \rho_c(p_{in})$ so that for any $\rho<\rho_c$ clusters cannot be detected. In Figure~\ref{f1b} we plot the detection boundaries on the $(p_{in}, \rho)$ plane for different values of $\a+\g$.  For each $\a+\g$, the corresponding boundary separates two regimes, so that points above  (above) the separator correspond to detectable (undetectable) clusters. 

\section{Clustering with hard constraints}
\label{sec:hard}
We now consider the case of  hard constraints, by setting $w=\infty$~\footnote{Note that fixing $w$ to some large but finite number of $O(N)$ will guarantee that all the constraints are satisfied.}. In this case, the cavity equation involves all the order parameters, which makes its analysis more complicated. Instead, we address this case by solving the cavity equation using {\rm population dynamics}. 

\begin{figure}[!htb]
\centering
\subfigure[]{
    \includegraphics[width = 0.45\textwidth ]{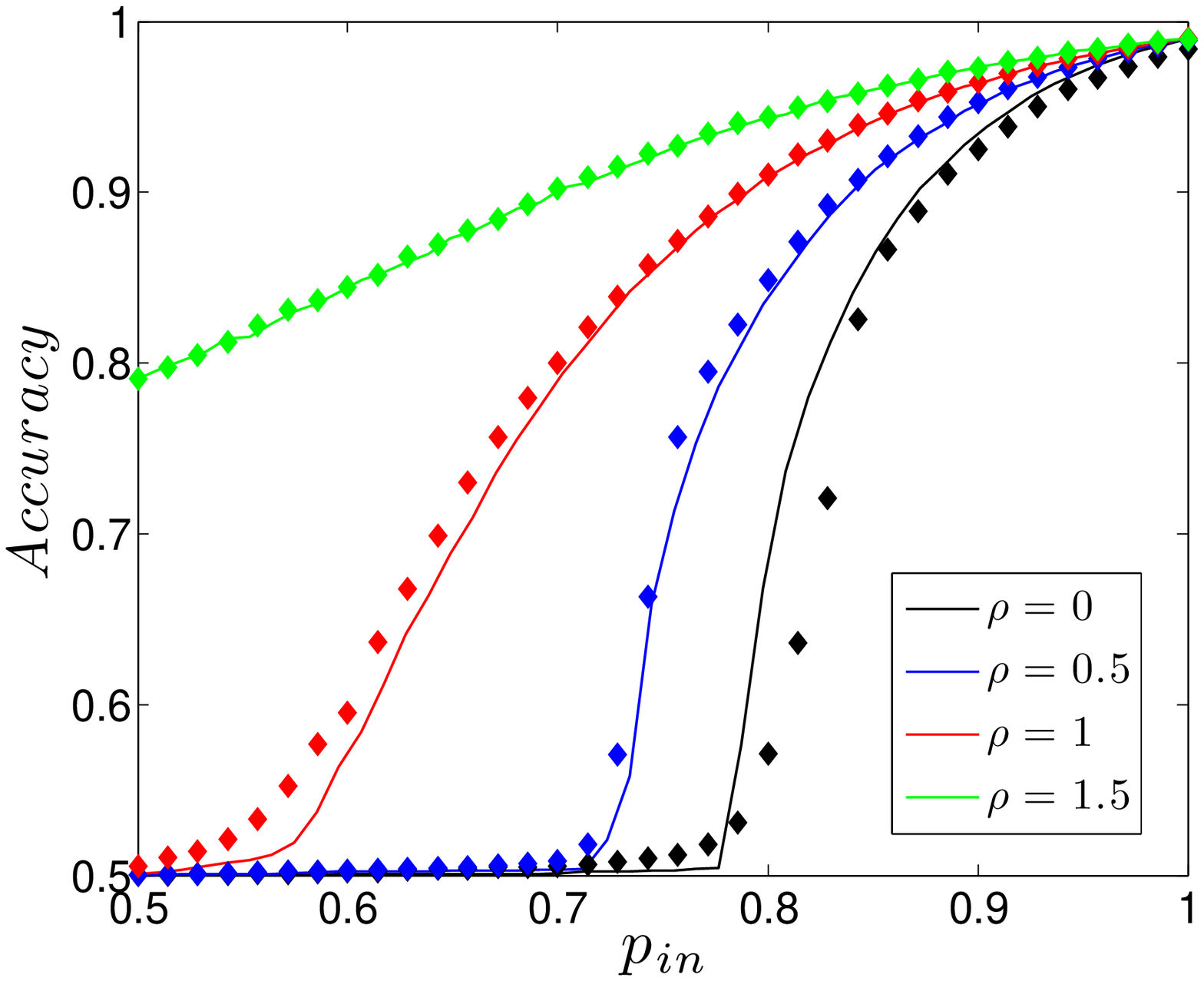} \label{f2a}
    }
    \subfigure[]{
    \includegraphics[width = 0.45\textwidth]{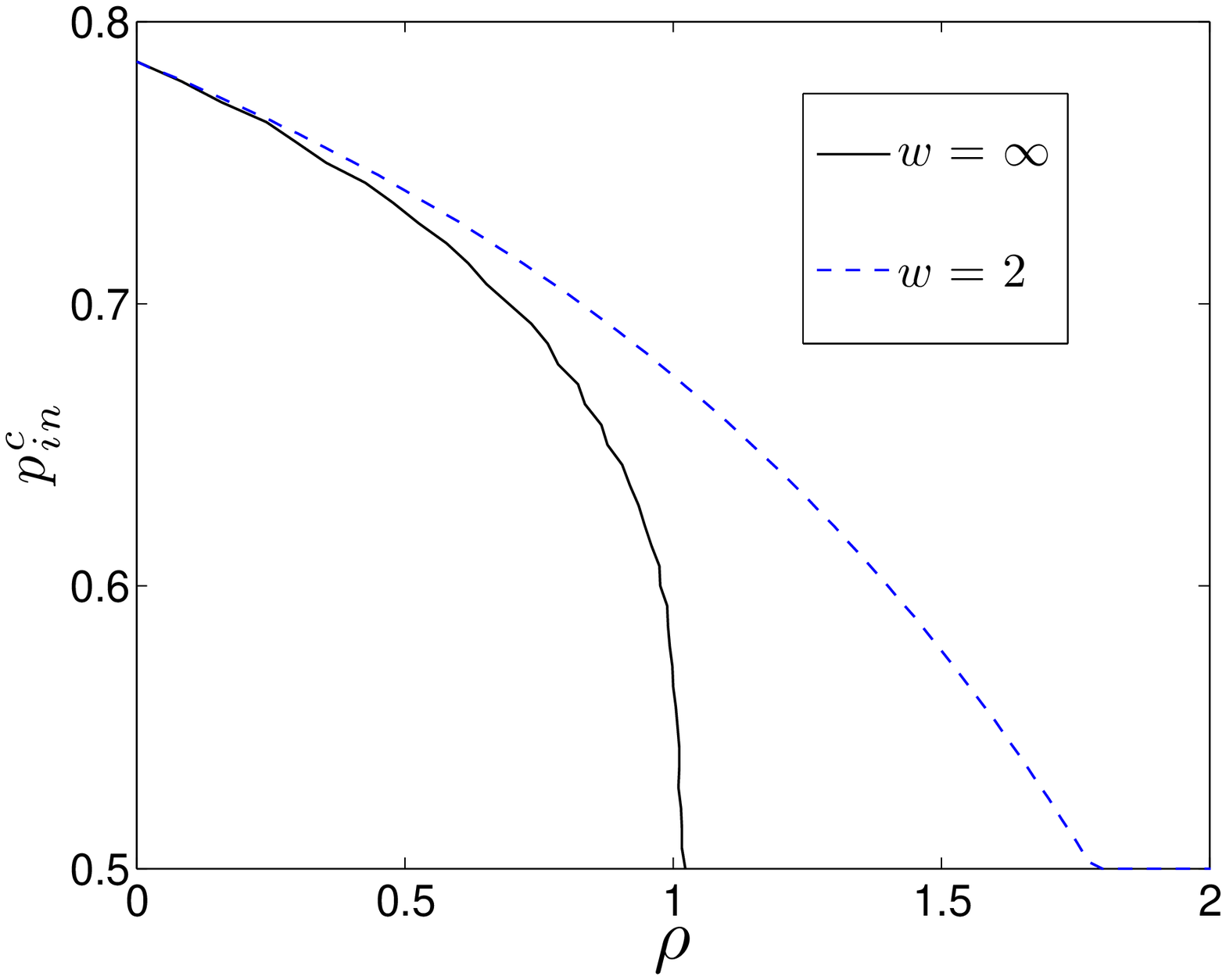} \label{f2b}
    }
        \caption{(a) Magnetization plotted against the mixinf parameter $p_{in}$ for $\a+\g=4$ and different $\rho$.  Lines are generated from population dynamics and points are generated from simulated annealing. From bottom to top we have $\rho = 0,0.5,1,1.5$. (b) Dependence of the critical mixing parameter $p_{in}^c$ on $\rho$ for $\a+\g=4$. In contrast to the  $w=2$ case (dashed line), the detection threshold for $w=\infty$ attains its minimum value $p_{in}=1/2$ when $\rho$ is slightly greater than $1$.}
\end{figure}

We begin with a brief description of population dynamics which is described in more detail in an identical setting in \cite{mezard4}. The goal is to estimate the distribution of fields $P(h)$ whose fixed point is Eq.~\ref{cavity}. One starts with a large population of $\mathcal{N}$ random fields, $h_i$, then generates a new random cavity field, $h_0 = \ssum_{k=1}^N \phi[h_k,J_{0k}] - \ssum_{k=1}^N \phi[g_k,K_{0k}] $, where $J_{0k},K_{0k}$ are drawn according to some fixed distribution and the other fields, $h_k, g_k$ are drawn randomly from the population. We then replace a random field in the population with our newly calculated cavity field. The stationary distribution of this Markov chain corresponds to the fixed point of Eq.~\ref{cavity} for large populations. We can then calculate the magnetization using Eq.~\ref{arto}, approximating $P(h) \sim \sum_{i=1}^{\mathcal{N}} \delta_{h_i,h} / \mathcal{N}$. We used $\mathcal{N} \sim 100,000$ for results depicted in Fig.~\ref{f2a},\ref{f2b}. 
We also compare our results to simulations using synthetic data. After generating random graphs of size $N=100,000$, we find the ground state of the Hamiltonian \ref{ham} using simulated annealing. 

For a subset of nodes connected by labeled edges, we can determine the relative group membership for any pair in the group due to the transitivity of the constraints. Therefore, finding node assignments that satisfy hard link constraints on a graph amounts to a two-coloring problem and can be done efficiently. As we add random edges to a graph, the size of the connected clusters is well-known~\cite{erdosrenyi}. For $\rho<1$, most clusters are disconnected and the size of the largest cluster is $O(\log(N))$. At $\rho=1$, we reach the ``percolation threshold'' where the size of the largest cluster goes like $O(N^{2/3})$. Once $\rho>1$, $O(N)$ nodes belong to one giant connected component. We will investigate the consequences of these different regimes below.

Looking at the results of Fig.~\ref{f2a}, we see that for small amounts of supervision, $\rho < 1$, the impact of the constraints is to shift the detection threshold to smaller values of $p_{in}$. Qualitatively, this is no different than the effect of adding more unlabeled edges within clusters. This behavior is expected, since adding hard constraints is equivalent to studying the same unsupervised clustering problem on a  {\em renormalized} graph (e.g., merging two nodes that are connected via constraints). This is in contrast to results for prior information on nodes in~\cite{armen}, which showed that even small amounts of node supervision shifted the detection threshold to its lowest possible value $p_{in}=1/2$. 

As $\rho \rightarrow 1$, there is a qualitative change in our ability to detect clusters.  A large number of nodes, $O(N^{2/3}),$ are connected by labeled edges. If we take the relative labeling of nodes in this largest group as the ``correct'' one, than we have a situation similar to node supervision, which, as discussed, moves the detection threshold to $\alpha=\gamma$.
While this large labeled component suffices to create non-zero magnetization in finite graphs (as seen from the simulated annealing results), as $N$ gets large, the effect of this component diminishes.
For $\rho > 1$, we see that the fraction of nodes contained in the largest labeled component suffice to produce non-zero magnetization even at the group-defining threshold $\alpha=\gamma$. 


We define the detection threshold $\rho_c$ as the minimum value of $\rho$ so that $\rho > \rho_c \rightarrow m(\alpha,\gamma,\rho) > 0$.
In Fig.~\ref{f2b}, we investigate the location of the detection threshold $\rho_c$ for a graph with fixed connectivity $\alpha+\gamma = 4$, as a function of the cluster overlap, $p$. For $p \sim 1/2$, where there is little cluster structure, we see that $\rho_c$ is just greater than 1. As noted, this is the regime where a fraction of nodes are fixed by edge constraints. As $p \rightarrow 1$ and the cluster structure becomes more well-defined, the necessary number of labeled edges per node falls below $1$, where the size of clusters connected by edge constraints is only logarithmic in the number of nodes. Finally, $\rho_c$ approaches $0$ at $p \approx 0.786$. This $p$ is the point at which the magnetization is non-zero, even without any supervision.

\section{Discussion}
\label{sec:discussion}

We have presented a statistical mechanics analysis of semi-supervised clustering in sparse graphs in the presence of pair--wise constraints on node cluster assignments. 
Our results show that addition of constraints does not provide automatic improvement over the unsupervised case. This is in sharp contrast with the semi--supervised clustering scenario considered in~\cite{armen}, where any generic fraction of labeled nodes  improves clustering accuracy.

When the cost of violating constraints is finite, the only impact of adding pair--wise constraints is {\em lowering} the detection boundary. Thus, whether adding constraints is beneficial or not depends on the network parameters (see Figure~\ref{f1a}). For the semi--supervising clustering with hard pair--wise constraints, the situation is similar if the number of added constraints is sufficiently small. Namely, for small density of constraints the subgraph induced by the must-- and cannot links consists mostly of isolated small components, and the only impact of the added constraints is to lower the detection boundary. The situation changes drastically when the constraint density reaches the percolation threshold. Due to transitivity of constraints,  this induces a non--vanishing subset of nodes (transitive closure) that belong to the same cluster, a scenario that is similar to one studied in Ref.~\cite{armen}. In this case, the detection boundary disappears for any $\a,\g$.  

In the study presented here, we assume that the edges are labeled randomly. One can ask whether other, non--random edge--selection strategies will lead to better results. Intuition tells us that the answer is affirmative. Indeed,  in the random case one needs to add $\rho=1$ additional edges per node in order to have the benefit of transitivity. For a given $\rho$, a much better strategy would be to choose $\rho N+1$ random nodes (rather than edges), and connect them into a chain  using labeled edges. This would guarantee the existence of a finite fraction of labeled nodes for any $\rho$.  

Finally, it is possible to envision a situation where one has access to two types of information -- about cluster assignment of specific nodes~\cite{armen}, and pairwise constraints such as studied in the present paper. Furthermore, this information might be available at a cost that, generally speaking, will be different for either type of information.  An interesting problem then is to find an optimal allocation of a limited budget  to achieve the best possible clustering accuracy. 

\acknowledgments

This research was supported in part by the National Science Foundation grant
No. 0916534, US ARO MURI grant No. W911NF0610094, US AFOSR MURI grant
No. FA9550-10-1-0569, and US DTRA grant HDTRA1-10-1-0086. Computation for the work described in this paper was supported by the University of Southern California Center for High-Performance Computing and Communications (www.usc.edu/hpcc).
%


\end{document}